# Wide bandwidth instantaneous RF spectrum analyzer based on nitrogen vacancy centers in diamond

M. Chipaux,[1,a)] L. Toraille,[1,b)] C. Larat,[1] L. Morvan,[1] S. Pezzagna,[2] J. Meijer[2] and T. Debuisschert[1]

[1]*Thales Research & Technology, 1 Av. Augustin Fresnel, 91767 Palaiseau Cedex, France*

[2]*Department of Nuclear Solid State Physics, Institute for Experimental Physics II, Universität Leipzig,*

*Linnéstr. 5, D-04103 Leipzig, Germany*

We propose an original analog method to perform instantaneous and quantitative spectral analysis of microwave signals. An ensemble of nitrogen-vacancy (NV) centers held in a diamond plate is pumped by a 532 nm laser. Its photoluminescence is imaged through an optical microscope and monitored by a digital camera. The microwave signal is converted to an oscillating magnetic field in the area of the NV centers by a loop shaped antenna. Induced magnetic resonances are detected through a decrease of the NV centers photoluminescence. A magnetic field gradient induces a Zeeman shift of the resonances and transforms the frequency information into spatial information, which allows for the simultaneous analysis of the microwave signal in the entire frequency bandwidth of the device. The time dependent spectral analysis of an amplitude modulated microwave signal is demonstrated over a bandwidth of 600 MHz, associated to a frequency resolution of 7 MHz and a refresh rate of 4 ms. With such integration time, a field of a few hundreds of µW can be detected. Since the optical properties of NV centers can be maintained even in high magnetic field, we estimate that an optimized device could allow frequency analysis in a range of 30 GHz, only limited by the amplitude of the magnetic field gradient. In addition, an increase of the NV centers quantity could lead both to an increase of the microwave sensitivity and to a decrease of the minimum refresh rate down to a few µs.

---

a) Present address : Department of Biomedical Engineering, University of Groningen / University Medical Center Groningen, Antonius Deusinglaan 1, 9713 AV Groningen – The Netherlands
b) Present address : Ecole Normale Supérieure de Lyon, 15 parvis René Descartes - BP 7000 69342 Lyon Cedex 07 – France

The interception and the study of the quickly changing radio-frequency signals generated by the most recent communication and sensing systems require performing time-frequency analysis of the largest possible bandwidth, ideally a few tens of GHz. The most direct method would consist in directly digitizing the signal of interest and to apply standard digital filtering methods, such as time-overlapped Fourier transforms. However, due to the limited performances of high-speed digitizers and to the large computational resources required by the processing, this approach is usually limited to an instantaneous bandwidth of a few hundreds of MHz only. When combined with a heterodyne down-conversion setup and a tunable microwave local oscillator, it may analyze over a few tens of GHz, but with a low probability of intercept. In the past decade, analog alternatives have been investigated[1–3]. Avoiding both direct sampling and heterodyning, they perform a real-time spectral analysis over a wide instantaneous bandwidth of a few tens of GHz and provide a 100 % probability of intercept of the signals. However, they either lack of frequency resolution[1], or require very low temperature operation[2,3].

In this paper we propose an original alternative based on ensembles of nitrogen vacancy (NV) centers in diamond. We demonstrate the instantaneous acquisition of the full spectrum over a 600 MHz bandwidth with 7 MHz resolution, in a compact and simple setup operating at room temperature. This first demonstration can be extended to potentially cover a bandwidth of a few tens of GHz. The structure of the paper is as follows. We first present the NV center in diamond and recall its main properties that lead to the detection of microwave signals. In a second part we present and study the experimental set-up. Then the spectral analysis of a time dependent amplitude modulated signal in the microwave range is demonstrated. Finally we propose an estimation of the theoretical limits of this technique.

The NV center is a point defect of diamond constituted of a nitrogen atom (N) and a vacancy (V) in two adjacent crystallographic sites. Its main axis, denoted $(N-V)$, joins the nitrogen atom to the vacancy and follows one of the four crystallographic directions $(a)$, $(b)$, $(c)$ and $(d)$ of the diamond lattice. The NV defect is a perfectly photostable color center. When pumped with green light (at 532 nm in our case), it emits a red photoluminescence signal (from 600 to 800 nm). It is an atom-like system that possesses an internal degree of freedom linked to its electronic spin that can be both initialized and readout through optical excitations[4]. Two electronic spin resonances, induced by an oscillating magnetic field in the microwave range can be detected by a decrease of the NV center photoluminescence. Applying an external magnetic field $\vec{B}$ induces a Zeeman shift that lifts the degeneracy of those resonances. When the transverse part of the magnetic field is negligible with respect to the longitudinal part, the Zeeman shift is linear and proportional to $B_{\parallel}$, the projection of the field along the $(N-V)$ axis[5]. The frequencies $\nu_-$ and $\nu_+$ of the resonances are given by:

$$\nu_{\pm} = \nu_0 \pm \gamma_{NV} \cdot B_{/\!/} \tag{1}$$

Where $\nu_0 = 2.87$ GHz is the frequency of the resonances with no external magnetic field and $\gamma_{NV} = 28$ MHz/mT is the electron spin gyromagnetic ratio of the NV center.

Taking profit of those properties, the NV center has been intensively studied as a quantitative magnetic field sensor[6]. Eq. (1) is used to determine $B_{/\!/}$ measuring the frequencies of the magnetic resonances. A first approach is to position a nanodiamond including a single NV center at the end of the tip of an atomic force microscope. Magnetic cartography with sensitivities down to 300 nT/$\sqrt{\text{Hz}}$ associated to a resolution of 10 nm have already been reported[7]. An alternative is to use a high density layer of NV centers close to the surface of a macroscopic diamond plate. A standard optical microscope can thus provide vectorial images of the magnetic field with no requirement of any scanning procedure[8–11].

The original technique described here works the reverse way. An ensemble of NV centers in a bulk diamond is used to directly sense the microwave signal. A static magnetic field, with controlled spatial distribution, associates a Zeeman shift, and thus a resonance frequency, to each position in the sample. Then, when a microwave signal with an unknown frequency is applied, a magnetic resonance appears at the position corresponding to that frequency. It is thus possible to reconstruct the whole frequency spectrum of a complex signal in one exposure of the camera. Since no data processing is required, the acquisition is instantaneous for the entire bandwidth, which makes possible the real-time spectral survey of quickly changing signals with a high probability of intercept.

Measurements were done with a single crystal diamond plate from Element Six grown by chemical vapor deposition. Its lateral dimensions are 4.5 $x$ 4.5 mm and the thickness is 500 µm. The two main faces are (100) oriented and the four facets are (110). The sample holds around 1 ppm of nitrogen impurities. A part of them is naturally converted to NV centers along the four crystallographic axes of the diamond.

A 532 nm wavelength gaussian pump beam is sent into the diamond plate through one of its facet (FIG. 1(a). The beam waist is 50 µm, corresponding to a Rayleigh range of 1.5 cm, which ensures a well collimated beam in its entire path through the crystal. The induced photoluminescence is collected by a ten times magnification optical microscope objective and sent to a digital camera (uEye UI-5240CP) through a lens with a 75 mm focal length (FIG. 1(b)). The total magnification of the microscope (around 4.3) is matched to the size of the active area ($\approx$ 1 mm).

The diamond is submitted to a magnetic field distribution oriented along the $(x)$ axis. It is produced by four parallelepipedal magnets ($6 \times 5 \times 2$ mm) positioned in the corners of a 8 mm side square (FIG. 1(c)). Due to this configuration, the magnetic field and its gradient (around 25 mT/mm) remain aligned all along the medium axis of the square. The diamond is then inclined by 35 ° in order to ensure a good alignment of the crystallographic direction $(a)$ of the diamond with the magnetic field (FIG. 1(d)). Finally, a loop shape antenna at the end of a coaxial cable converts the microwave electric signal inside the wire to an oscillating magnetic field in the area of the NV center. The induced magnetic resonances are locally detected by a relative drop of the photoluminescence image. Their frequencies are given by the position along $(x)$ axis. The contrast $C$ of a resonance is defined by:

$$C = \frac{S_0 - S}{S_0} \qquad (2)$$

where $S$ and $S_0$ are respectively the photoluminescence signal with and without magnetic resonance. It quantifies the strength of the resonance.

Although the antenna was not optimized for that particular application, we are able to detect monochromatic signals of a few hundreds of μW with an integration time of 4 ms. For such low power signals, the contrast C of the resonance has been measured to be proportional to its intensity. This is consistent with the standard electrodynamic model of the NV center[12]. This could be used as a calibration to extract quantitative data on the intensity. This also shows that our system can give quantitative data on the spectral components of the incoming signal provided an optimized antenna is designed.

The resonance distribution is characterized using a controlled microwave frequency sweep from 1.88 to 3.12 GHz. The normalized photoluminescence $C$ is recorded and plotted in FIG. 1(e) as a function of the position on the $(x)$ axis and the frequency. Several dark lines corresponding to the magnetic resonances are visible on the picture. They can be attributed by pairs to the four crystallographic axes of the diamond. Note that since our microwave synthesizer is limited to 3.2 GHz, the high frequency branches of the resonance lines are not visible on the figure. Two anticrossings are visible on FIG. 1(e). They correspond to the zero-field splitting of the NV center where the projection of the magnetic field along the NV centers axis is zero. In our case, the measured frequency is 2.93 GHz instead of the usual 2.87 GHz value. This is attributed to a residual transverse magnetic field that induces an upward shift of the upper energy levels and a downward shift of the lower energy level, which induces an increase of the transition frequencies[5]. Since our microwave synthesizer is limited to 3.2 GHz this shift of the frequency transition allows a larger useful frequency range, and we have thus chosen to work in that configuration.

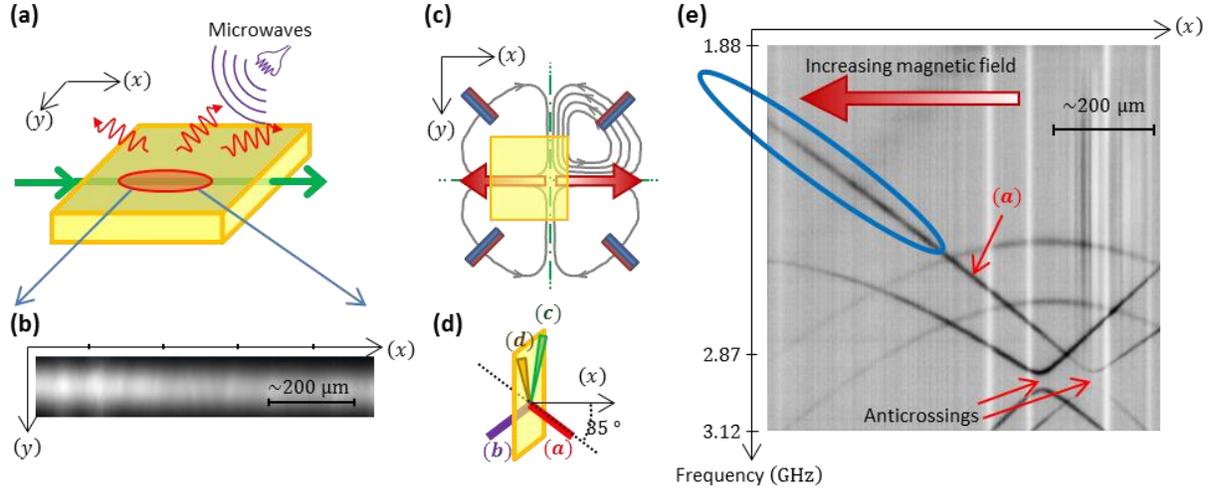

**FIG. 1 (a)** A CVD diamond plate (yellow) holds NV centers in its volume. It is pumped by a laser beam at 532 nm (green arrow). The induced red photoluminescence is collected through a standard optical microscope (not represented here) and monitored by a digital camera. **(b)** typical image of the monitored photoluminescence. **(c)** The diamond is submitted to a magnetic field gradient along the $(x)$ direction generated by an assembly of four magnets. The horizontal green dashed line indicates the plane in which the magnetic field is included. In particular, the field is aligned with the red arrow and is zero in the center of the square. In this area, the magnetic gradient is also aligned with that direction. **(d)** The four crystallographic axes $(a)$, $(b)$, $(c)$ and $(d)$ are represented relatively to the (110) facets of the diamond plate. The magnetic field is aligned with $(a)$, with an angle of 35° with axis $(x)$. **(e)** The normalized photoluminescence, obtained from the raw images, is plotted as a function of the position along the $(x)$ axis and of the microwave frequency. ESR lines are detected by a drop of the photoluminescence. Their frequencies are linked to the projection of the magnetic field along the crystallographic axes $(a)$, $(b)$, $(c)$ and $(d)$. The spectra represented in figure 2 were performed in the area delimited by the blue oval.

If we now move from the anticrossings to the left side of the figure where the magnetic field is high, we can observe a nonlinear Zeeman shift for three of the pairs. This also can be attributed to a non-negligible transverse component of the magnetic field[5] with respect to the crystallographic axes. For one of the pairs, the Zeeman shift is mostly linear, which indicates that the magnetic field is well aligned with the $(N - V)$ axis, here referred as direction $(a)$. The effect of the slight residual magnetic field is to induce a mixing of the spin states which results in a small decrease of the photoluminescence contrast[5], as can be seen on FIG. 1(e). Despite the complexity of the figure, we can select an area where only one resonance $(a)$ is present. There, a bijection is established between the resonance frequency and the position along the $(x)$ axis, delimited by the blue oval on FIG. 1(e). The maximum frequency range we obtained goes from 1.9 to 2.5 GHz.

A first demonstration is performed measuring the frequency spectrum of a simple amplitude modulated (AM) signal. It is generated multiplying a carrier signal at 2.2 GHz by a modulation signal at 40 MHz. This AM signal is first characterized by a conventional spectrum analyzer. Its two side bands, at 2.16 GHz and 2.24 GHz, have a power around 0 dBm, and the carrier at 2.2 Ghz is around −10 dBm. This signal is then sent to the antenna to induce the magnetic resonance signal of the NV centers. The normalized photoluminescence $C$ is plotted in FIG. 2(a) as a function of the spatial axes $(x)$ and $(y)$. The reference is the photoluminescence measured in the absence of any microwave excitation. The carrier and the two side-bands induce three resonances that are detected by a decrease of luminescence. Their frequencies are linearly linked to the position along the $(x)$ axis. However, a more accurate calibration can be done by measuring the exact position of the ESR line for each frequency of the spectrum. The resulting spectrum is plotted in Fig. 2(a). The linewidth of the lines is approximately 7 MHz, defining the spectral resolution of the device. With respect to the 600 MHz bandwidth, it corresponds to 85 channels. Here the two side-bands of the incoming signal have equal power. The observed difference of contrast corresponds to the non-uniformity observed in FIG. 1(e). It can thus be attributed to the slight misalignment of the magnetic field along $(a)$. One can also note that the resonances have a visible dependence in $(y)$ which is a confirmation of this slight misalignment. However, the corresponding frequency shift remains smaller than the resonance linewidth.

We then demonstrate the ability of our device to monitor instantaneous frequency variations. The AM signal is now time-dependent. The frequencies of both carrier and modulating signal are swept while the relative photoluminescence is monitored by the camera. Taking advantage of the low dependence in the direction $(y)$, the signal-to-noise ratio is increased

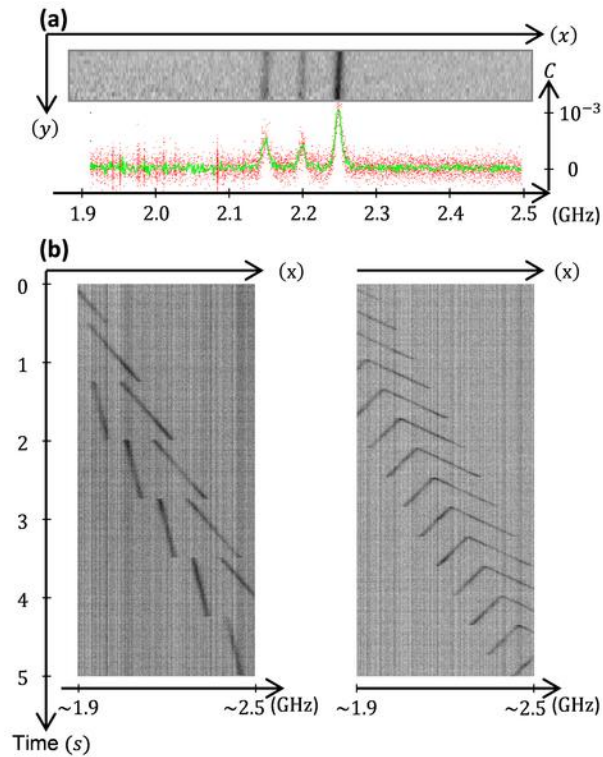

**FIG. 2 a)** The relative photoluminescence image of the NV centers submitted to an AM signal is plotted in grey levels along $(x)$ and $(y)$ directions. The carrier at **2.2 GHz** and the two side-bands at **2.16 GHz** and **2.24 GHz** are visible. The spectrum (red dots) is obtained plotting the relative photoluminescence of each pixel and exploiting the correspondence between position and frequency. The green line corresponds to a resampling with a **5 MHz** resolution that allows eliminating the high frequency noise of the raw data.
**b)** Spectrograms: The average relative photoluminescence is displayed in a 2D plot with one axis representing the frequency $\nu$, and the other one representing the time. The carrier is swept from **1.8** to **2.6 GHz** in **5 s**. The modulated signal is swept from **40** to **80 MHz** in **0.7 s** (left) and from **0** to **100 MHz** in **0.4 s** (right).

by summing the ESR signal column by column. The result corresponds to the instantaneous spectrum of the input signal. The acquisitions are repeated in time to monitor the temporal evolution of this spectrum. Resulting spectrograms are plotted in FIG. 2(b). Those curves show the versatility of our device as an instantaneous spectrum analyzer.

One limitation of the monochromatic linewidth is the hyperfine interaction between the electron spin of the NV center and the nuclear spin on the nitrogen atom. The $^{14}$N, which is the most abundant isotope of the nitrogen (99.64 %), leads to three resonance lines separated by 2 MHz[13]. The combination of the power broadening[14], produced by the continuous excitation of the NV centers, and of the decoherence caused by the electron spin bath of nitrogen impurities[6] induces a broadening that makes the hyperfine interaction difficult to resolve. To overcome this, one could take profit of diamonds of higher purity[15].

The total bandwidth is first limited by the presence of the eight resonances that correspond to the four crystallographic direction of the diamond. The use of preferentially oriented NV centers obtained in CVD growth in direction $(111)$[16,17] and $(113)$[18] could strongly attenuate the three unwanted resonances, and release the constraints on the selection of a range with bijective correspondence between frequency and position. Another advantage of such diamonds would be to increase by four the contrast of the single remaining resonance ($a$). The range is also limited by the size of the interaction area (limited in our case to the 1 mm diameter of the antenna) compared to the strength of the magnetic gradient. Both of them could be improved by a more optimized design. Typically, a magnet can provide a magnetic field of about 1 T that corresponds to a range of 30 GHz and therefore to more than 4,000 channels. As demonstrated in reference 5, the magnetic resonance properties of NV centers are maintained in this entire range provided the magnetic field is precisely aligned with the $(N-V)$ axis. Thus, such a broad microwave bandwidth should then be accessible with our technique.

Finally, the refresh rate corresponds to a reasonable exposure time of the camera regarding the amount of photoluminescence that is collected from the NV centers. The use of a diamond crystal with a higher NV concentration and of a microscope objective with higher aperture would allow a smaller exposure time and thus an increase of this refresh rate. A lower limit to the time resolution will be of a few microseconds, given by the photophysics of the NV centers. Indeed the complete photoluminescence cycle involves a metastable state which lifetime is around a few hundreds of nanoseconds[19]. This creates a time laps in which the NV center is not active, limiting the interrogation rate. Fortunately, this might not decrease the intercept probability of quickly changing incoming signals since many NV centers are available simultaneously.

In summary, the instantaneous spectral analysis of a quickly changing microwave signal with an ensemble of NV centers has been demonstrated over a bandwidth of 600 MHz, associated to a frequency resolution of 7 MHz and to a refresh rate of 4 ms. Those results can possibly be extended to a bandwidth of a few tens of GHz and the refresh rate can be decreased to a few microseconds.

The research has been partially funded by the European Community's Seventh Framework Programme (FP7/2007-2013) under the project DIADEMS (grant agreement n° 611143) and by the French Agence Nationale de la Recherche (ANR) under the project ADVICE (grant ANR-2011-BS04-021).